\documentclass[prb]{revtex4}% Physical Review B

\usepackage{graphicx}% Include figure files
\usepackage{dcolumn}% Align table columns on decimal point
\usepackage{bm}% bold math

\begin{document}

%\preprint{APS/123-QED}

\title{Geoneutrinos and the Earth inner parts structure}
% Force line breaks with \\

\author{V.V. Sinev}
% \altaffiliation[Also at ]{Physics Department, XYZ University.}%Lines break automatically or can be forced with \\
\affiliation{%
Institute for Nuclear Research RAS, Moscow\\
%This line break forced with \textbackslash\textbackslash
}%

\date{\today}
% It is always \today, today,
%  but any date may be explicitly specified

\begin{abstract}
The connection between geoneutrino registration and the Earth theory test is discussed. 
We compare standard theory of lithosphere plates and hypothesis of hydride Earth. Last hypothesis adds additional neutrino source $–$ 
planet core in which the initial Earth composition is conserved. Large volume scintillation detector is supposed to install at Baksan neutrino 
observatory INR RAS at Caucasus. The detector will register all possible neutrino fluxes, but mainly geo-neutrinos. So kind a detector (or detector net) 
placed in a number of sites on the Earth surface can measure all radioactivity from $^{238}$U and $^{232}$Th, because their neutrino energy 
exceeds the inverse beta-decay reaction threshold. By this way it will it possible to understand if there are any more neutrino sources in the Earth 
other than the crust and mantle.
\end{abstract}

%\pacs{Valid PACS appear here}% PACS, the Physics and Astronomy
                             % Classification Scheme.
%\keywords{Suggested keywords}%Use showkeys class option if keyword
                              %display desired
\maketitle

\section*{Introduction}

In 2005 in the Nature journal the KamLAND Collaboration has reported about observing antineutrinos from the Earth, so-called geoneutrinos [1]. 
Recently there was similar message from Collaboration BOREXINO [2]. Both measurements are within predictions of BSE [3] and Reference 
Model [4].

The geoneutrinos registration is important mainly when looking for the sources for Earth thermal flux. Total Earth thermal 
flux now is estimated 
as 44$\pm$1 TW and at least a half of it may be explained by radioactive elements placed in the inner parts according to
Reference Model.

One can put the question in the other way, where inside the Earth radioactive isotopes are concentrated, and what it may mean? 
This is similar to using isotopes, as a diagnostic tool in medicine, to test geophysical theories that offer different areas of the Earth, 
where radioactive elements can be located. The standard modern theory, for example, rejects the possibility of occurring uranium, 
thorium and potassium in the Earth's core. One could check this by using the antineutrino radiation emitted by beta radioactive nuclei 
belonging to the families of $^{238}$U and $^{232}$Th. Antineutrinos freely pass through the Earth, carrying out information about its source to the 
observer on the surface.

Antineutrinos emitted by nuclides from families of $^{238}$U and $^{232}$Th can be detected through the inverse beta-decay reaction 
on proton because their energy exceeds the threshold of this reaction. This reaction is widely used to study antineutrino radiation 
from reactors. It is perfectly studied theoretically and has the largest cross section between other possible reactions of neutrino interactions with matter. 

There exist also the hypothesis of natural nuclear reactor occurring in the center or at the junction of the core and the mantle [5], 
which is closely related to geoneutrinos. This hypothesis provides an opportunity to explain the missing amount of heat, the source 
for Earth's magnetic field and the periodic change of the magnetic poles of the Earth. It can also be verified by recording neutrinos from the Earth.

To register the natural fluxes of antineutrinos and neutrinos at low energies ($<$100$-$150 MeV) a large scintillation neutrino spectrometer 
is proposed. They plan to install it at Baksan Neutrino Observatory (BNO) INR RAN at a depth of 4800 m.w.e. Planned spectrometer 
target mass is about 5 kt. 

A number of studies also proved the possibility of using such a detector are under consideration for the moment [6-8]. 

\section{Generally accepted theory of the Earth structure}

Our days the widely accepted theory of the Earth is the theory of slabs or lithosphere plates with various modifications. 
According to these ideas Earth crust consists of slabs floating on the surface of the molten mantle, sometimes coming closer to 
each other, sometimes moving off. 

Inside the Earth consists of a series of nested layers [9], whose boundaries were determined by seismic methods. 
The top layer $-$ the crust. The boundaries of the crust bottom are determined quite accurately, since at the Earth's 
surface there placed a lot of seismic stations that monitor seismic activity of Earth. Based on these data there exist a map 
of inner layers of the crust with step of $2\times2^{\circ}$ [10]. The thickness of the crust varies from 5-8 km on the ocean floor up to 30-60 km 
under the continents. 

Below the crust ending with Mohorovich layer begins upper mantle, extending to a depth of 660-670 km. In composition, this part 
of mantle consists of silicates and oxides of metals such as mainly Si, Mg, Fe, and Al.

The upper mantle is separated from the lower mantle with a thin, presumably, olivine layer. Lower mantle reaches the Earth's core. 
The composition of this part of mantle is not clearly defined. Probably it consists of some metal alloys. 

The core, as well as the mantle is divided into two parts: the liquid outer core (2900 km to 5150 km) and the solid inner one 
deeper than 5150 km. 
The core consists mainly of iron with additions of Ni and some other metals. 

The temperature of the earth according to various estimations reaches 4000-6000$^{\circ}$ C in the center. 

The boundaries of the inner layers of the Earth are shown in Fig. 1.

\begin{figure}
\includegraphics{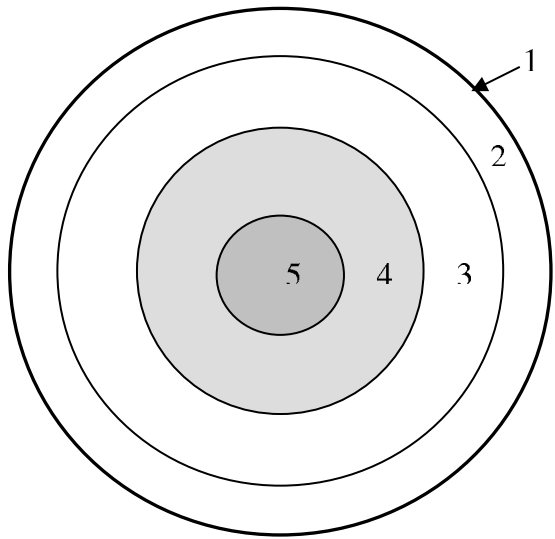}% Here is how to import EPS art
\caption{\label{fig:epsart1} The boundaries of the inner layers of the Earth according to the seismological measurements. 
1 $-$ Earth's crust, 2 $-$ upper mantle, 3 $-$ lower mantle, 4 $-$ outer core (liquid), 5 $-$ the inner core (solid).}
\end{figure}

\section{Alternative theory}

There are also other theories of the Earth constitution that can be called as alternative. One of them was suggested by Russian geologist 
V.N. Larin in 1968 [11]. The boarders of inner layers of the Earth are the same, they were measured experimentally by 
registering seismic waves propagation in the Earth. The main difference in theories is in the composition of the layers. 

According to Larin's theory the Earth has a slightly different composition than the conventional theory. 
He proposed the composition of Earth inner parts on base of elements distribution analysis of the solar system. 
It was found that there is a dependence of elements abundances on the distance from the Sun [12, 13]. This dependence is 
explained by the hypothesis of F. Hoyle of elements separation in the magnetic field of the protostar in accordance with the ionization potential. 
Fig. 2 shows the dependence of elements abundances from the degree of ionization in the Earth-Sun system, the Earth-Moon and the 
Earth-meteorite belt. The figures imply that the elements in the primary solar system might be separated according to the ionization 
potential by the protosun magnetic field (see Table. 1). 

On the basis of this hypothesis the original composition of the Earth was proposed, which is given in Table 1.

\begin{figure}
\includegraphics{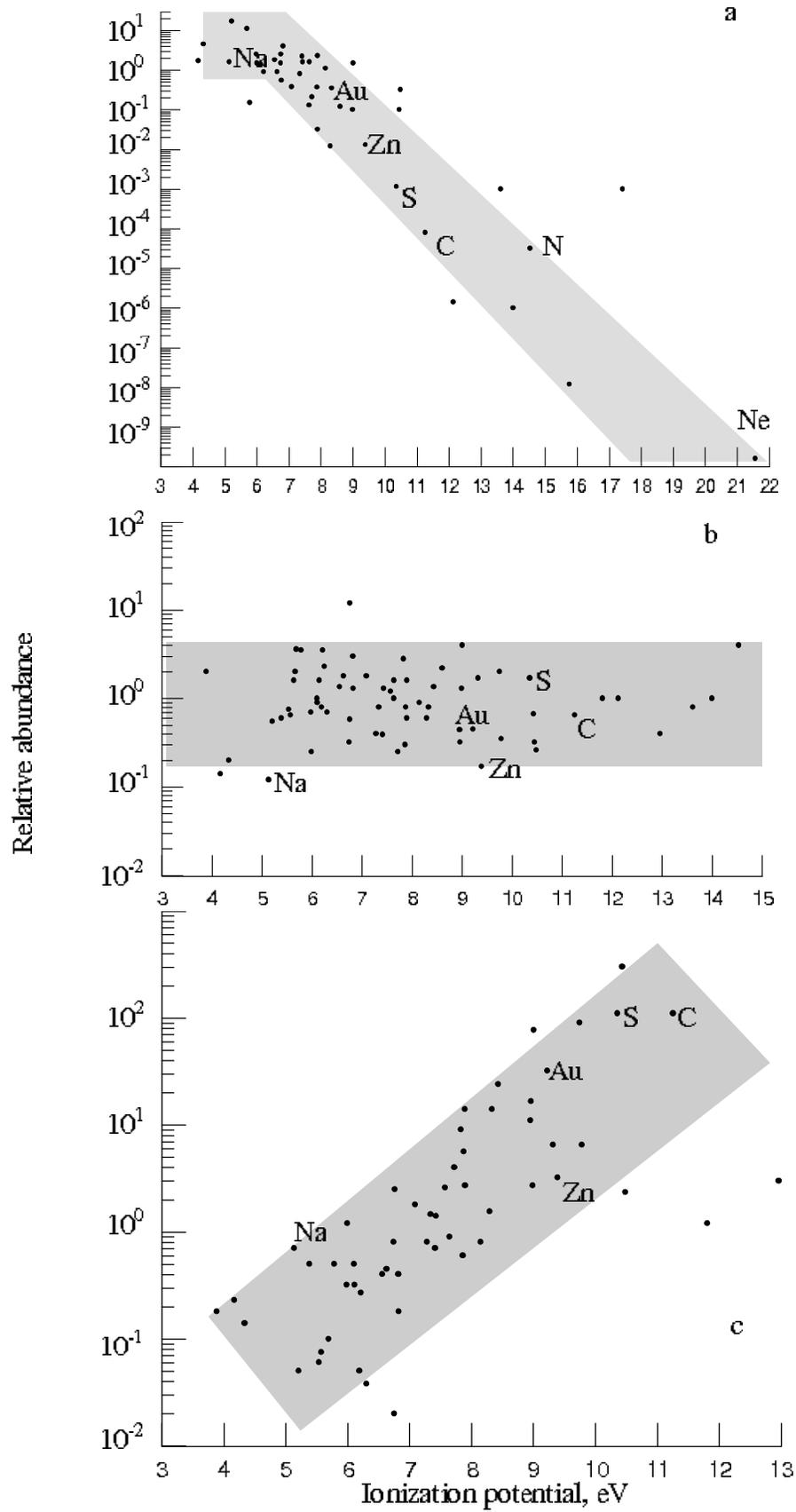}% Here is how to import EPS art
\caption{\label{fig:epsart2} The dependence of the element abundance from the ionization potential. (a) $-$ Earth-Sun, (b) $-$ Earth-moon system, 
(c) $-$ the Earth-meteorite belt.}
\end{figure}

\begin{table}[h]
\caption{The primary composition of the Earth (based on magnetic separation)}
\label{table:1}
\vspace{10pt}
\begin{tabular}{l|c|c}
\hline
Elements  & Atomic \% & weight \% \\
\hline
Si & 19.5 & 45 \\
Mg & 15.5 & 31 \\
Fe & 2.5 & 12 \\
Ca & 0.9 & 3 \\
Al & 1.0 & 2 \\
Na & 0.7 & 1.5 \\
O & 0.6 & 1 \\
C & 0.03-0.3 & 0.03-0.3 \\
S & 0.01-0.1 & 0.03-0.3 \\
N & $<$0.01 & $<$0.01 \\
H & 59 & 4.5 \\
\hline
\end{tabular}\\[2pt]
\end{table}

From Table 1 follows the conclusion that the Earth can not contain in the centre iron core. According to this theory hard core consists 
of light metal hydrides such as Si, Mg, Al and Fe. Liquid core consists of the same metals, but saturated with hydrogen. 
Experiments have shown that metals can be liquid at relatively low temperatures, but high pressures under the condition of 
saturation with hydrogen (see [11]). 

According to Larin the Earth has not liquid mantle. Instead of it there is a mixture of metals and silicides, i.e. compounds of 
metals with silicon in so-called mantle. And instead of the upper mantle is the shell of oxides and silicates. There are inclusions 
of metals saturated with hydrogen in the liquid phase. The crust is formed mainly by oxides of metals. 

The main conclusion of his theory that the Earth inside is relatively cold. All the energy of gravitational compression went 
on the hydrides production. 

Earth's heat comes from the decay of radioactive elements that were originally located on all the thickness of the Earth. 
This heat causes the disintegration of hydrides and the additional heat output. Escaping hydrogen flows out this heat to the surface 
with some metals and radioactive elements, falling into liquid phase. Hydrogen rising to the surface assembles into the jets. 
On the surface at these locations may exist orogenesis and volcanic activity. Additional heat occurs near the surface when going up 
silicides and hydrogen meet with oxygen and atmospheric water in the gaps of crust rocks.

\section{Geoneutrino}

Let's regard radioactive elements occurring inside the Earth. There are mostly $^{238}$U, $^{232}$Th and $^{40}$K. 
We do not account now some other isotopes like $^{87}$Rb and $^{235}$U because of their small amount and weak input to 
total thermal flux. These isotopes produce heat by means of alpha and beta decays. 
When the beta-decay occurs there appeared electron antineutrino simultaneously with beta electron. 
Antineutrinos easily escape from the Earth boarders and go out, they can be detected with some detector placed on the surface. 
These antineutrinos frequently are called geoneutrinos. 

Theories presented behigh place radioactive isotopes in different reservoirs. According to modern ideas practically all
radioactive elements are placed in the crust and in the mantle in equal proportions. In opinion of geochemists and 
geophysicists the core can not contain radioactivity. But Larin's theory states of their existance in the core in enough 
quantities. The core is conserved the primordial content of the Earth corrected on decay time. Some isotopes do not exist more,
for example $^{237}$Np which halftime is 2.2$\times$10$^6$ years.

Antineutrino energy spectra of named elements are well known. At figure 3 one can see calculated energy spectrum of 
geoneutrinos originating from $^{238}$U and $^{232}$Th. They demonstrated here because their energy exceeds the threshold
of inverse beta-decay reaction that is widely used for antineutrinos registration. It will be described below.

We can estimate the effect produced by the antineutrinos in a detector if to do some suppositions about the location of sources. 
In table 2 one finds some calculations of antineutrino effects at possible detector positions. Calculations are based on the 
distribution of radioactive elements in crust and mantle and data on the crust depth from [10]. Last column contains our 
calculation supposing $^{238}$U and $^{232}$Th existance in the core. 

\begin{figure}
\includegraphics{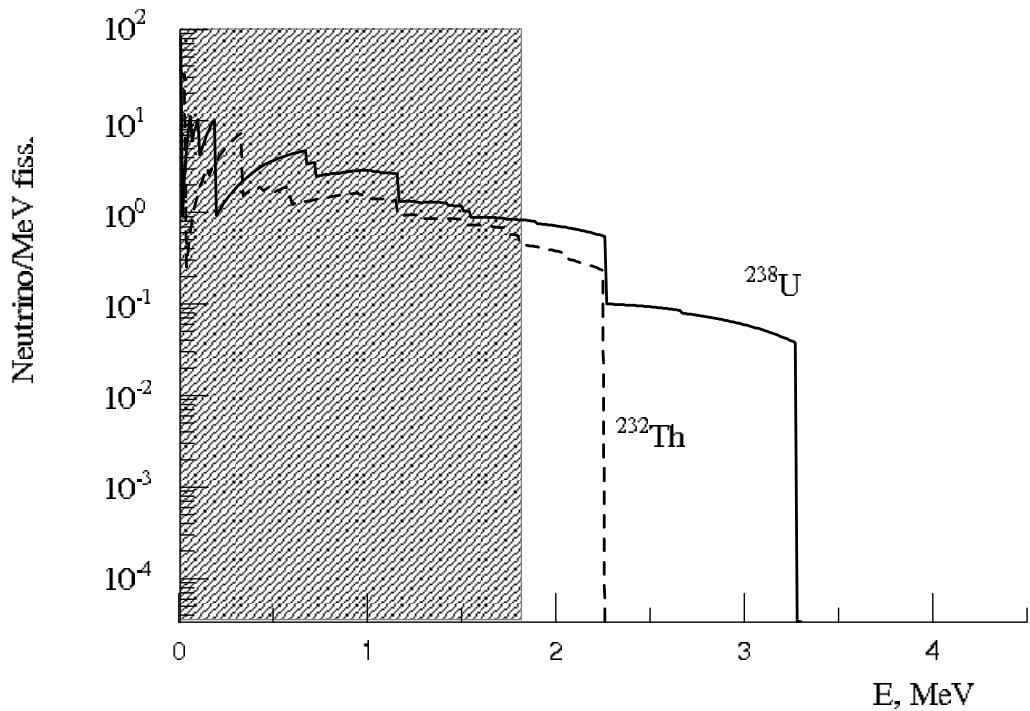}% Here is how to import EPS art
\caption{\label{fig:epsart3} The spectra of antineutrinos from $^{238}$U, $^{232}$Th. The shaded area is inaccessible to measurement
because it lays below the threshold of reaction (1).}
\end{figure}

\begin{table}[h]
\caption{Geoneutrino effect calculations for some detector locations}
\label{table:2}
\vspace{10pt}
\begin{tabular}{l|c|c|c|c|c}
\hline
Location  & [3] & [14] & Crust up to the sea level & Total crust thickness & With the Earth core \\
\hline
Hawaii & 12.5 & 13.4 & 15.99 & 16.03 & 20.8 \\
Kamioka & 34.8 & 36.5 & 33.2 & 33.4 & 38.2 \\
Gran Sasso & 40.5 & 43.1 & 41.7 & 42.4 & 47.1 \\
Sudbury & 49.6 & 50.4 & 52.2 & 52.8 & 57.5 \\
Pyhasalmi & 52.4 & 52.4 & 55.4 & 55.7 & 60.5 \\
Baksan & 51.9 & 55.0 & 55.1 & 57.0 & 61.5 \\
\hline
\end{tabular}\\[2pt]
\end{table}

The column "Total crust thickness" accounts the part of the crust higher than the sea level, it increases a little bit 
detector counting rate in places surrounding with mountains. The core affects the same way all detectors on the surface because 
the distance from it to there is the same.

\section{Detectors and detector projects for geoneutrinos}

Neutrino detector, located on the surface of the Earth can register neutrino flux from its interior. 
Antineutrino can be registered through the inverse beta decay reaction on proton which has the largest cross section between other
neutrino interactions
\begin{equation}
\bar{\nu_{e}}+p \rightarrow n + e^{+}.
\end{equation}

The positron appeared as a result of the reaction carries out practically all antineutrino energy. 
Its kinetic energy is linearly connected with antineutrino energy
\begin{equation}
T=E-\Delta -r_n,
\end{equation}
where $T$ -  positron  kinetic energy, $E$ -  antineutrino energy, $\Delta$ - the reaction threshold equals to 1.806 MeV and $r_n$ is neutron recoil energy.

So, the positron spectrum is the same as antineutrino's, but shifted on 1.8 MeV and convoluted with cross section. 
Recoil energy in the first approximation can be neglected.

In a detetor placed on the surface the antineutrinos from crust will be registered in greater amount than from the mantle if
the crust is enough thick. In the other hand, if the crust is thin, larger effect will be from the mantle.

In Japan and Italy our days they just measure geoneutrinos effect. Collaboration KamLAND has began measurements in 2002 and in 2005 
reported about observing the signal from geoneutrinos [1]. Authors gave the value of 28$\pm$15 for geoneutrinos counting rate
what gives 57$\pm$32 TNU (1 TNU = 1 ev. per year per 10$^{32}$ protons on the target). 
This result is in accordance with Reference Model [4] predictions.

The leading constraints in this experiment is nonremovable background from nuclear power plants surrounding the detector, see 
table 3. At fig. 4 one can see measured positron spectra from reaction (1) in KamLAND detector from a number of sources 
including geoneutrinos.

Detector BOREXINO at Gran Sasso (Italy) has the effect from geoneutrinos compared with nuclear power plants 
effect but the value
of it is much smaller than one at KamLAND. They reported about 9.9$\pm$4 events observed during 252.6 ton$\cdot$years.

It is necessary to note that these detectors KamLAND and BOREXINO were constructed for other goals than geoneutrinos 
registration and the fact of discovering the Earth antineutrino emission by them says about very high sensitivity of 
these both detectors.

To understand better the problems of the Earth thermal balans, search of the input of $^{238}$U and $^{232}$Th in total 
Earth produced heat, containment of uranium and thorium un the core and mantle etc. we need to have sufficient statistics. 
That is why the firther movement in research of geoneutrinos is connected with construction of more powerful detectors, 
the same class as KamLAND but situated at far distance of nuclear power plants [16]. Proposed in INR RAS detector belongs
to so kind of detectors. Its counting rate as expected are to be about 220 events per year.

Detector construction accounts advantages of KamLAND and BOREXINO but has more massive target to supply higher statistics 
that can be accumulated in observable time. The schetch of detector is shown at fig. 6. We have chosen target mass 5 kt as
a step to more massive detector. In the other hand it can be regarded as a part of complex detector. Complex detector may be 
a detector summarising all working detectors of the similar type (BOREXINO + KamLAND + SNO+ + etc.).

\begin{figure}
\includegraphics{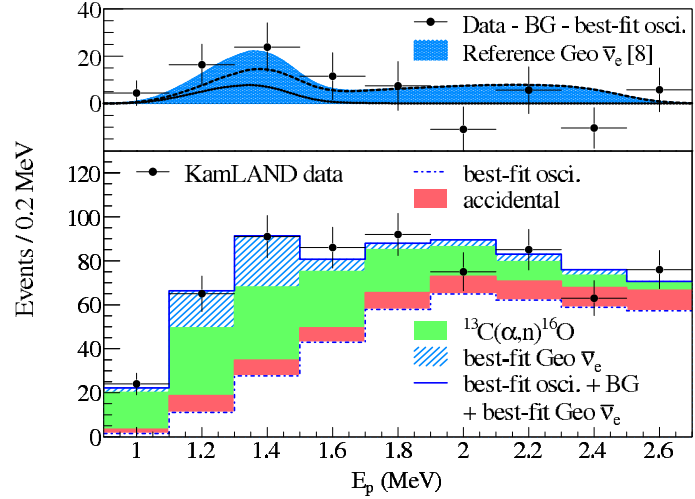}% Here is how to import EPS art
\caption{\label{fig:eps4} Positron spectra from reaction (1) in the detector KamLAND.}
\end{figure}

\begin{figure}
\includegraphics{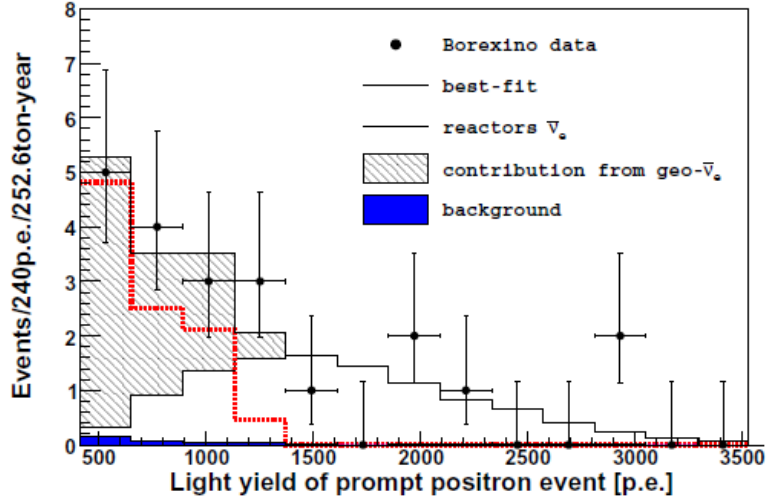}% Here is how to import EPS art
\caption{\label{fig:eps5} Positron spectra from reaction (1) in the detector BOREXINO.}
\end{figure}

\begin{figure}
\includegraphics{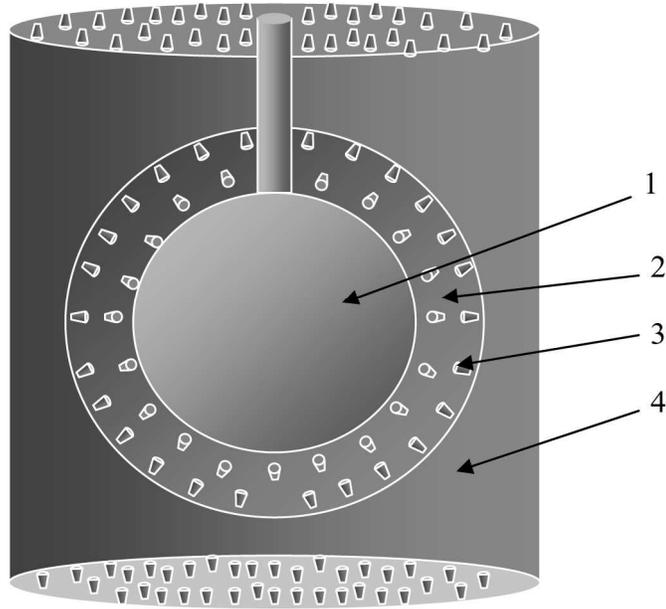}% Here is how to import EPS art
\caption{\label{fig:eps6} Schematic view of the detector, proposed to be installed in the Baksan Observatory INR RAS. 
1 $-$ target 5 kt of liquid scintillator, 2 $-$ passive protection from natural radioactivity, 3 $-$ PMT, 4 $-$ Region of anticoincidence.}
\end{figure}

Observable positron spectrum from geoneutrinos in Baksan detector is shown at fig. 7.

There are also other projects of large volume detectors for geoneutrinos. One of them is project LENA - Low Energy Neutrino 
Astronomy {17].
The project is going to be installed in Pyhasalmi (Finland). It has a goal to make a target of 50 or 90 kt of liquid 
scintillator. The scintillator is supposed to be viewed by 12 thousand PMTs. The volume is needed here to supply also the
experiment for looking for proton decay.

\begin{figure}
\includegraphics{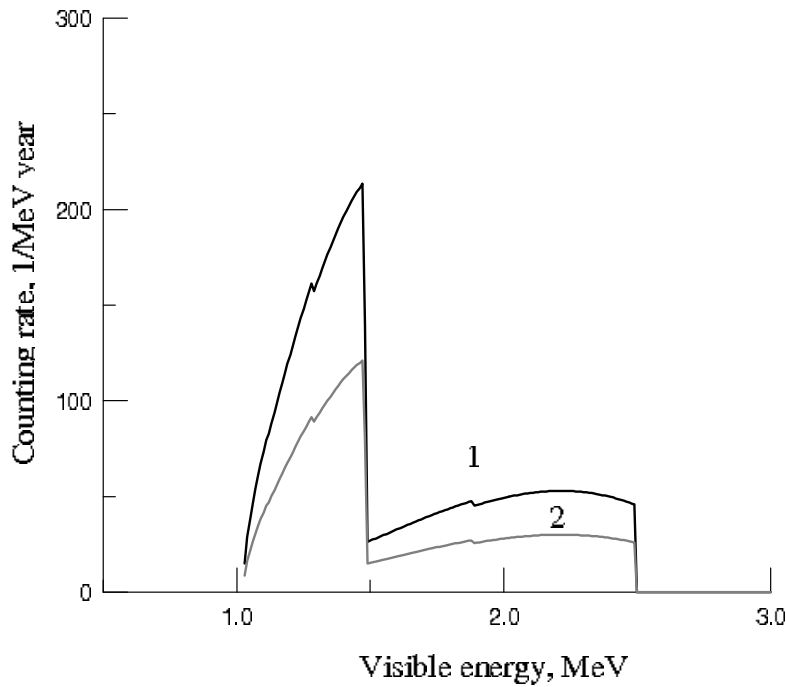}% Here is how to import EPS art
\caption{\label{fig:epsart7} Positron spectra from geoneutrinos in the detector at BAKSAN neutrino observatory INR RAS.
1 $-$ spectrum without oscillations, $\sim$108 ev./year; 2 $-$ spectrum accounting oscillations, $\sim$61.5 ev./year. 
Calculations were made for a target containing 10$^{32}$ protons.}
\end{figure}

Another project of a huge detector is proposed by the University of Hawaii, this is project Hano-Hano [18]. The detector
is placed on the ship and can be transported to the choosen position by the sea. At the point it will be sunk to the bottom
at a depth of $\sim$4000-5000 meters. This detector will search mainly mantle geoneutrinos because the crust at the ocean 
bottom is very thin. This detector is very attractive because of very low background of NPPs, see table 3. The target mass is 
expected to be 10 kt.

In the nearest future one more large volume detector is going to be ready. The project SNO+ at Sudbury (Canada) will use 
1 kt liquid scintillator for geoneutrinos registration [19]. This project is based on the previous detector SNO that was used 
for solar neutrinos detection with a target of 1 kt of heavy water. Now they supposed to change heavy water on liquid 
scintillator.

We could note that Baksan project [16] has some advantages. It is placed enough far from power plants and background from 
reactors will be the smallest excluding Hano-Hano project. The crust at Caucasus is enough thick and counting rate will be
hopefully high. Also this site has just ready infrastructure and qualificated manpower.

\section{Determination of neutrino direction}

When measuring the neutrino direction we can obtain more information on the neutrino sources position. In some works there
was shown a possibility to get the neutrino direction through the analysis of angle distribution of reaction (1) 
products [6, 15].

At low energies neutron goes exactly in the same direction as antineutrino [19, 20], but positron is emitted almost 
isotropically slightly back. During the moderation and diffusion the point of neutron capture on hydrogen or gadolinium
is shifted averagely forward in the neutrino direction. In practice one can measure the distance between positron and neutron
registration places and obtain angle distribution.

In [7] they have shown that the average neutron displacement distance depends on the neutrino sources positons in the Earth.
Neutrinos from the crust give small displacement ($\sim$ 0.29 cm according Monte Carlo simulation), 
from lower mantle $-$ 1.2 cm. We can 
write the expression for displacement as a function of a low mantle fraction in total neutrino flux.
\begin{equation}
d_z=(1-{\alpha}_m)d_{cr}+{\alpha}_m d_m,
\end{equation}
where $d_{cr}$ and $d_m$ are displacements for the crust and the mantle, 0.29 and 1.2 cm relatevely. 
${\alpha}_m=F_m/(F_m+F_cr)$ - the lower mantle fraction of total neutrino flux. At fig. 8 one can see this dependence. 
To say something defined one needs to have at least 4000 events, when the uncertainty of displacement measurement
becomes about 0.3 cm what is compareable with $d_{cr}$ [7]. And observing at this statistics a displacement larger
than 0.3 cm will signify the existance of radioactive elements in lower mantle or in the core. There may be done calculations 
also for the core as for the mantle.

\begin{figure}
\includegraphics{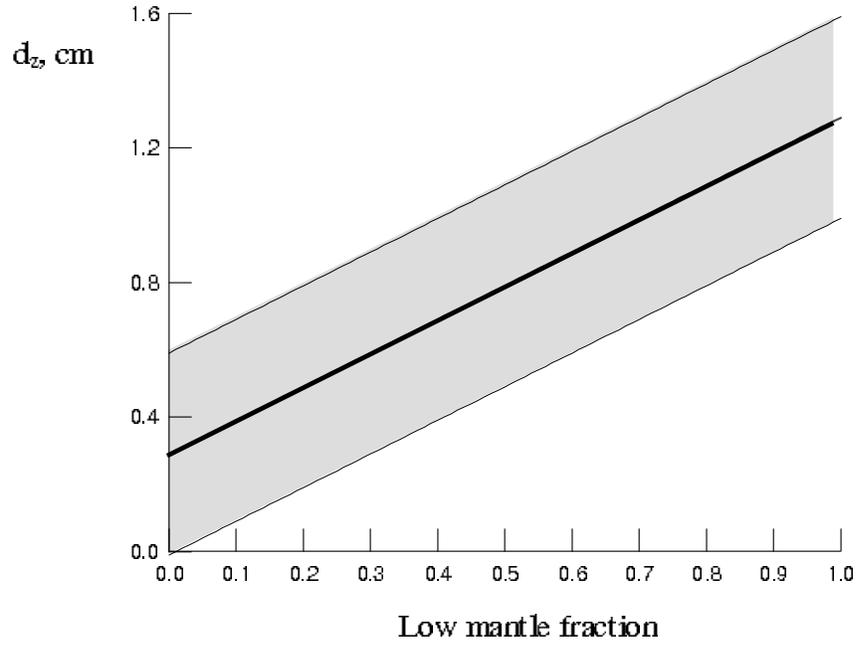}% Here is how to import EPS art
\caption{\label{fig:epsart8} Average neutron shift from the point of birth as a function of radioactive elements
fraction in low mantle.}
\end{figure}

\section{Nuclear power plants background}

One of backgrounds for geoneutrino detector will be the antineutrino emission from powerful nuclear reactors built
all over the world. In most of sites it is compareable with geoneutrino effect. Fortunately it can be calculated with
an accuracy $\sim$3\% [7] and accounted in analysis.

At fig. 9 one can see calculated positron spectrum at Baksan site from all Nuclear power plants, there were taken into 
account 412 nuclear reactors. To estimate resulting spectrum we used standard antineutrino spectrum taken from [20].
At the spectrum there exist many peaks that can be explained by the geometry of the site and reactors. Individual reactor 
spectrum is distorted according to the reactor distance from the site.

For the Baksan site reactors counting rate is estimated to be about 160 events/year for total target mass 5 kt ($\sim$40
ev./year per kt, see fig. 9). In table 3 one can find reactor counting rates for other sites. Calculation is done for 
10$^{32}$ protons in a target for better comparison.

\begin{figure}
\includegraphics{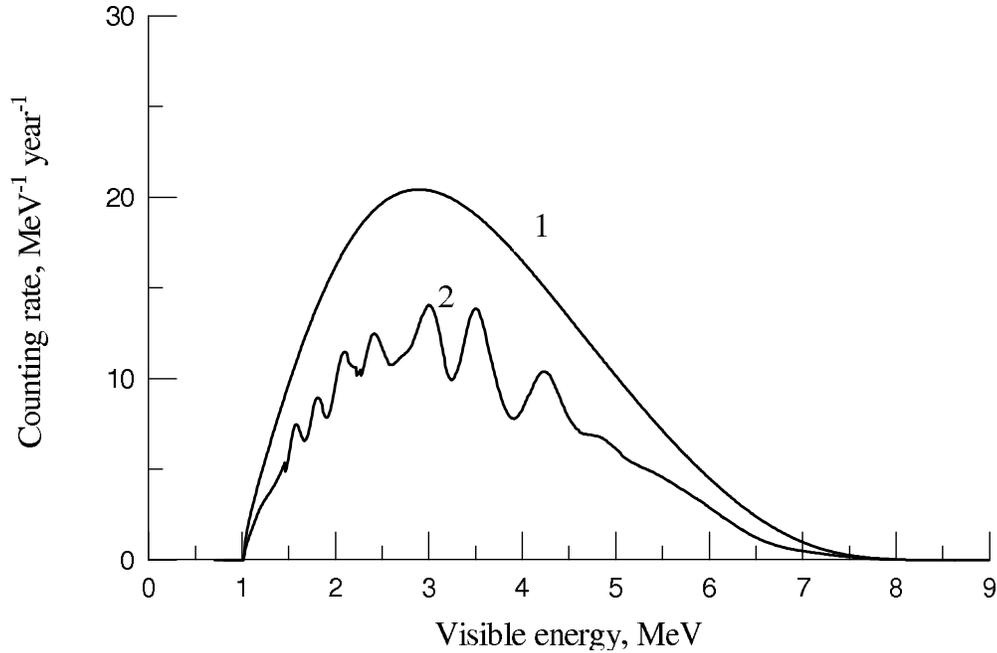}% Here is how to import EPS art
\caption{\label{fig:epsart9} Spectra of positrons observed in the detector from industrial reactors: 1 $-$ spectrum without oscillations (70 ev./year), 
2 $-$ spectrum accounting oscillations (40 ev./year). For 10$^{32}$ target protons.}
\end{figure}

\begin{figure}
\includegraphics{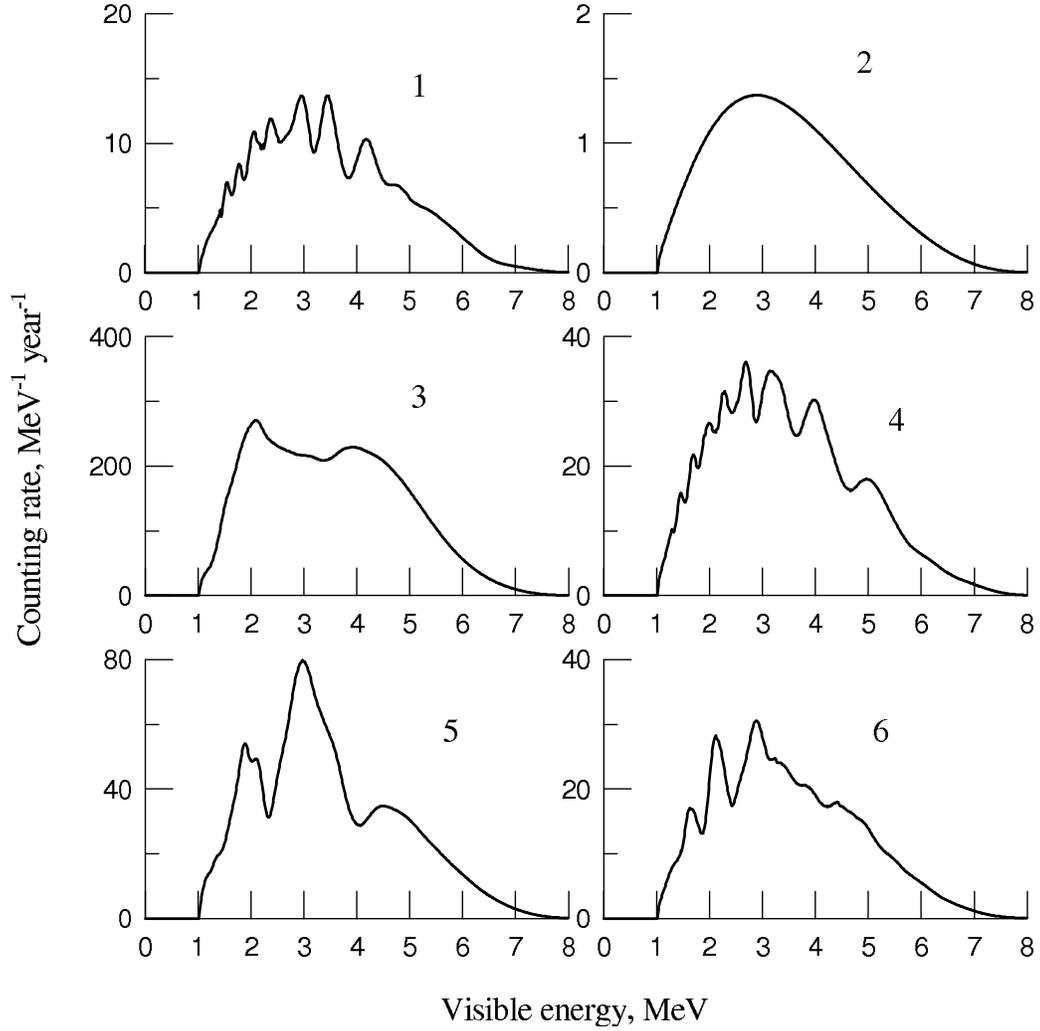}% Here is how to import EPS art
\caption{\label{fig:epsart10} The same as at fig. 9 for several detector sites: 1 $-$ Baksan ($\sim$40 TNU), 
2 $-$ Hawaii ($\sim$4.7 TNU), 3 $-$ Kamioka ($\sim$928 TNU),
4 $-$ Gran Sasso ($\sim$111 TNU), 5 $-$ Sudbury ($\sim$198 TNU), 6 $-$ Pyhasalmi ($\sim$87 TNU).}
\end{figure}

\begin{table}[h]
\caption{Nuclear reactors counting rate in places of detector location. For 10$^{32}$ protons.}
\label{table:3}
\vspace{10pt}
\begin{tabular}{l|c|c}
\hline
Location  & Events per year & $<$R$>$, km \\
\hline
Hawaii & 4.7 & 7483 \\
Kamioka & 928 & 227\\
Gran Sasso & 111 & 1189 \\
Sudbury & 198 & 667 \\
Pyhasalmi & 87 & 1245 \\
Baksan & 40 & 2003 \\
\hline
\end{tabular}\\[2pt]
\end{table}

At fig. 10 the positron spectrum shape is shown for a number of detector locations. It is seen that parameters of 
neutrino oscillations can be obtained from common analysis of positron spectrum shape.

We have done calculation accounting that all reactors work at full power. It is possible to compare data from KamLAND 
detector done at [1] with our calculation. In [1] was reported about 258 events were detected during 515 days
at average reactor power 0.6. Total efficeincy accounting all selection criteria was 0.898. 
\begin{equation}
928.3\times 0.461\times 0.6\times 0.898=230.5.
\end{equation}
Our calculation is within the experimental error of direct measurement at KamLAND.

\section{Conclusion}

The detection of geoneutrinos is a new tool to investigate the inner parts of the Earth. Neutrinos do not meet barriers when
come from inner parts of the Earth to the surface. Neutrino detector placed on the surface can register neutrinos coming 
from total Earth thickness. A number of neutrino detectors placed at locatoions where some part of the Earth can prevail on
others can help in understanding how many radioactivity is in the crust and mantle, or may be in the core.

If to analyse angle distribution of neutron and positron as products of reaction (1) one can better locate neutrino 
sources inside the Earth and resolve if there are neutrino sources in the core (nuclear reactor or radioactivity).

Power reactors background can be analysed to improve neutrino oscillations parameters.

Neutrino detector proposed for insatallation at Baksan neutrino observatory satisfy all criteria for so kind a detector. It can 
be involved in the net of detectors for investigation of the Earth inner parts and other goals available for large neutrino
detectors. Chosen target mass is not enough for sensitive looking for proton decay, but with other detectors it can be regarded 
as a part of comlex compound detector.

\section*{Acknowledgments }

Author thanks L.A. Mikaelyan for the interest to the work and useful consultations and G.V. Domogatsky for useful discussions 
and support.

\end{document}